\documentclass[a4paper]{article}
\usepackage[T1]{fontenc}
\usepackage{graphics}
\usepackage{rotating}

\makeatletter

\newcommand{\LyX}{L\kern-.1667em\lower.25em\hbox{Y}\kern-.125emX\@}

\makeatother

\begin{document}

\title{A Price Dynamics in Bandwidth Markets for Point-to-point Connections\footnote{ ToN paper 01-1009. Submitted Jan. 17, 2001 to IEEE Trans. on Networking, revised Feb. 8, 2001}}

\author{Lars Rasmusson\protect\( ^{1}\protect \)\\
Erik Aurell\protect\( ^{1,2}\protect \)\\
 1) Swedish Institute of Computer Science\\
Box 1263, SE-164 29 Kista, Sweden \\
2) Dept. of Mathematics, Stockholm University\\
SE-106 91 Stockholm, Sweden}

\date{February 8, 2001}

\maketitle
\begin{abstract}
We simulate a network of \( N \) routers and \( M \) network users making
concurrent point-to-point connections by buying and selling router capacity
from each other. The resources need to be acquired in complete sets, but there
is only one spot market for each router. In order to describe the internal dynamics
of the market, we model the observed prices by \( N \)-dimensional It\^o-processes.
Modeling using stochastic processes is novel in this context of describing interactions
between end-users in a system with shared resources, and allows a standard set
of mathematical tools to be applied. The derived models can also be used to
price contingent claims on network capacity and thus to price complex network
services such as quality of service levels, multicast, etc. 
\end{abstract}

\section{Introduction}

To be able to provide guaranteed quality of service, QoS, in a network, a user
needs to be able to reserve capacity, or 'bandwidth', in congested routers.
The reservation scheme should be efficient in the sense that one reservation
should not unnecessarily block other reservations, and it should not require
extensive negotiation. One scheme that fits these requirements is to trade router
capacity in spot markets. The assumption is that someone reserves capacity for
a connection by buying the capacity in the routers along the cheapest path between
the source and destination node. When the capacity is no longer needed, it is
sold to someone else. Increased demand increases price, so alternative paths,
if they exist, may become competitive, and the users will tend to move their
bandwidth usage away from congested routers.

We ultimately want to be able to price contingent claims on resources, formalized
as options or futures. The working hypothesis is that adding a suitable set
of such claims will improve the efficiency of the resource allocation, as prices
will better reflect all available information. Indeed, the purpose of contingent
claims on future value, such as options, futures, or more generally, derivative
securities, is to construct a market for trading the expected future value.
For the problem studied here, this means trading in expected future demand and
supply. One way to price derivatives is to use statistical modeling of the price
dynamics, since under suitable assumptions, which we will cover in a forthcoming
separate contribution, derivative prices are functions of current market prices
and the statistical model\cite{Black73}\cite{Avellaneda99}. This paper addresses
the necessary preliminary issue of how to estimate parameters in two stochastic
models from observed market prices, and also presents measures of the efficiency
of the market resource allocation scheme. These questions are also of independent
interest, as the efficiency is to be used to compare different market system
with each other, and with other allocation schemes. 

Using artificial markets for resource allocation in distributed systems dates
back to the mid 80ies, ranging over markets for storage capacity \cite{Kurose89},
CPU time \cite{Ferguson88} \cite{Waldspurger92}, and network capacity \cite{Kurose85}
\cite{Sairamesh95}. The emphasis has been on evaluating the efficiency of the
resource allocation, rather than understanding the resulting price dynamics.
More recent work stresses the \emph{agent} aspect, i.e. that the trading parties
are locally optimizing entities \cite{Faratin00}. Combinatorial markets, i.e.
trading of bundles of distinct resources is yet a relatively new area. Somewhat
related is the area of combinatorial auctions \cite{Rassenti82} \cite{Rothkopf98}
\cite{Sandholm99}. The use of derivatives for network admission control has
been used in \cite{Lazar98}.

Previous bandwidth market models usually only include a primary market, in which
end-users can buy and sell capacity only from the router owner. In the presented
model end-users trade amongst each other, i.e. the capacity is traded on secondary
markets. 

This paper is organized as follows: in section \ref{sec:method} we present
the model, of which a central ingredient is a detailed mechanism of price formation,
with very low overhead in user-user communication. In section \ref{sec:Results}
we model mathematically the resulting price processes. The main tools used are
stochastic differential equations (It\^o-processes), their associated Fokker-Planck
equations, and the stationary distribution of those. In section \ref{sec:Discussion}
we sum up an discuss our results.

\section{Method\label{sec:method}}

\subsection{Market model}

\begin{figure*}[ht]
{\par\centering \resizebox*{7cm}{!}{\includegraphics{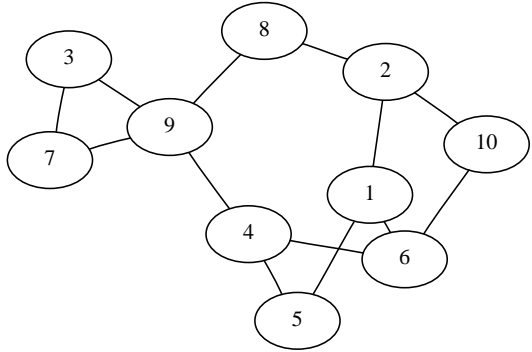}} \par}

\caption{\label{fig:network}The network with 10 routers.}
\end{figure*}
We simulate a network consisting of \( N \) inter-connected routers and \( M \)
network users concurrently making reservations of router capacity for point-to-point
connections. There is one spot market per router, in which users trade router
capacity. 

We use Farmer's non-equilibrium market dynamics \cite{Farmer00} as a prescription
of how prices change due to trading (see also sec. \ref{sec:discussionFarmer}).
Farmer's model is based on the assumption of a market maker that guarantees
liquidity at all times, and that buying causes prices to increase, while selling
causes prices to decrease. Farmer assumes that the price dynamics is such that
it is impossible to move the market by performing a sequence of trades, where
the net traded volume is zero, and that the relative price changes are independent
of the current value of the price. From these assumptions, Farmer derives a
formula that the price per unit in a transaction of \( \omega  \) units is
\( \tilde{S}(S,\omega )=Se^{\omega /\lambda } \), where \( S \) is the unit
price in the previous trade, and \( \lambda  \) the market depth or liquidity,
i.e. the rate at which the price is changed by trading. For a derivation, see
Appendix A. This model is useful since we do not have to simulate details of
the order-book in each market. Instead, we can calculate the price change caused
by trading directly from the last price and the size of the trade.

\subsection{Simulation setup}

\label{sec:setup}At each time interval, \( m \) new demands are generated.
A demand is a 7-tuple, 
\[
d\equiv <id,uid,src,dst,cap,dur,max>\]
 The user identities are chosen independently, so one user may receive zero
or one or several new demands. If the current time is \( t, \) demand \( id \)
specifies that the user \( uid \) demands \( cap \) units in each node on
a path from \( src \) to \( dst \), starting at time \( t \) and ending at
time \( t+dur \) if it costs less than \( max \) to obtain the resources.
If there are several paths, a choice will be made, see below. During the simulation
a user reserves capacity in a router by buying that capacity, and sells excess
capacity that is no longer in need. 

User \( i \) owns \( r_{i,j} \) units of \( j \). Initially, none of the
simulated users are assigned any resources or money, i.e. \( r_{i,j}=0,\, cash_{i}=0 \)
for all \( i \) and \( j \), nor do they have any resource demand, \( \omega _{i,j}=0 \).
When a user manages to satisfy a demand, its capital is increased by the amount
\( max \), and when resources are bought and sold, the capital is decreased
and increased, respectively, by the cost of the resource. The simulation is
run in \( L \) time steps from time \( 0 \) to \( T \) with time increments
\( \Delta t \). At each time step where the current time is \( t \):

\begin{itemize}
\item Generate \( m \) new demands. A demand is specified by a unique demand number
\( id \), a user \( uid \) randomly drawn from the set of \( M \) users,
a source node \( src \) and a destination node \( dst \), both randomly drawn
from the set of \( N \) nodes, and the required capacity per node \( cap \),
which is randomly drawn as the ceiling of \( e^{K\xi } \) where \( \xi  \)
is a uniform random variable between 0 and 1. If some demand is more important
than others, a user is willing to pay more for that resource. In this simulation,
\( max \) is a linear function of the required capacity, i.e. the maximum total
cost \( max=C_{unit}\, cap \), where \( C_{unit} \) is a simulation parameter.
The duration (number of time steps) \( dur \) is randomly drawn integer between
\( 1 \) and \( D \). 
\item Calculate \( \omega _{i,j} \), the net change of resource \( j \) of user
\( i=uid \) in the following way: 

\begin{itemize}
\item For each new demand \( d \), user \( i \) looks at the last known transaction
prices \( S_{j}(t) \) and decides to buy \( cap \) resources along \emph{the
least cost path.} Since the user will not know the actual cost of a buying and
later selling resources along a path, its decision on whether to buy resources
or not, is determined by a parameter \( C_{max} \). The user decides to buy
the resources if and only if the estimated total cost to buy the resources is
less than \( C_{max}max \). The resulting demand \( cap \) is then added to
the \( \omega _{i,j} \) for all the resources \( j \) on the least cost path,
and the amount \( max \) is added to \( cash_{i} \). 
\item For each satisfied demand \( d \) terminating at \( t \), decide to sell the
resources that were allocated (if any) to the demand, i.e. subtract the resulting
supply of liberated capacity on the least cost path if the demand was satisfied
and router capacity was bought. 
\end{itemize}
\item All the demands on capacity in single routers (i.e. \( \omega _{i,j} \)) are
effectuated one by one in random order, and for each trade prices are updated
to \( S_{j}e^{\omega _{i,j}/\lambda _{j}} \). The \( cash_{i} \) is decreased
by \( \omega _{i,j}S_{j} \) and the owned amount of resource is increased by
\( \omega _{i,j} \). According to the price formation formula, the trades are
made at prices which depend on the actual order. The prices payed by the users
are thus not the same as those used for determining the least cost path. The
next known price is the one that pertains after all the trades, and is independent
of the order. 
\item When all trading is done for this time-step, log the last transaction prices,
\( \hat{S}(t) \), the number of satisfied demands, and repeat.
\end{itemize}

\subsection{Simulation Parameters}

\label{sec:parameters}The simulation was run for \( L=1000 \) time-steps (\( \Delta t=0.01) \)
using the network in (fig. \ref{fig:network}). There were \( 10 \) routers
and 10 users, so \( N=10 \) and \( M=10 \). The liquidity in all Farmer markets
was chosen to be \( \lambda _{i}=10 \). The maximal cost per route users were
willing to accept was determined by \( C_{unit}=100 \) and \( C_{max}=1 \)
for all users. Initial prices were set to \( \hat{S}_{i}(0)=10 \) in all markets.
Every time step, \( m=10 \) new demands were generated. The duration \( dur \)
was uniformly distributed between \( 1 \) and \( 10 \), i.e. \( D=10 \).
The call duration sets the time scale in the simulation, as we will see below.
The required amount of capacity was determined by \( K=2 \) (see above).

\section{Results\label{sec:Results}}

\begin{figure*}[ht]
{\centering \begin{tabular}{cccc}
\begin{sideways}
\hspace{2cm} price
\end{sideways}&
\includegraphics{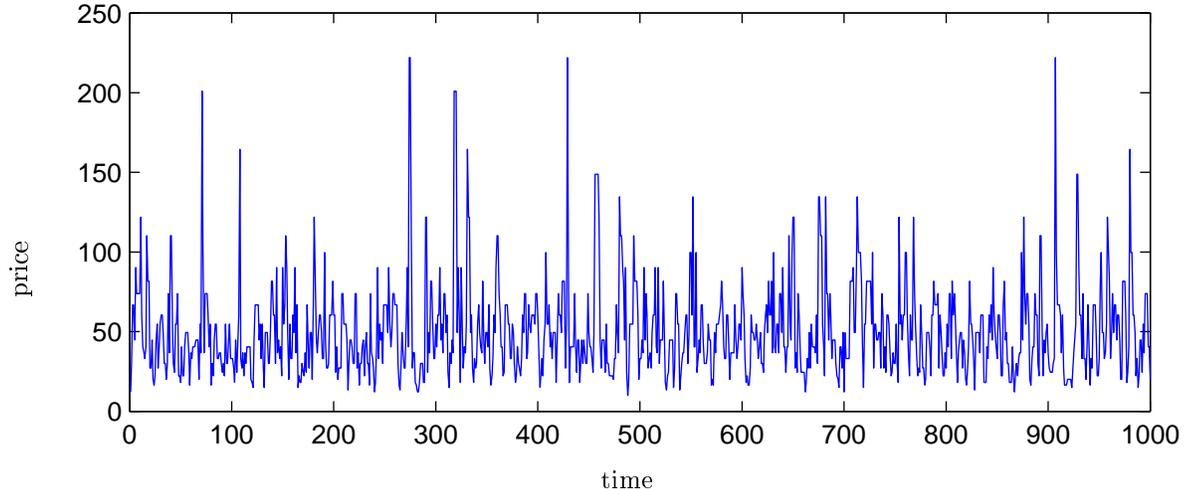} &
&
\\
&
time&
&
\\
\end{tabular}\par}

\caption{\label{priceRouter1}Plot of the unit price of capacity in router 1 from time
0.01 to 10. \protect\( \Delta t\protect \)=0.01.}
\end{figure*}

\subsection{Statistical modeling of the price process}

\paragraph{Inhomogeneous drift, additive noise, constant coefficients }

\label{sec:sde1}The price as shown in (fig. \ref{priceRouter1}) does not appear
to be drifting freely. Instead it appears to return towards the same area. Assuming
the price process is an Ornstein-Uhlenbeck process, the dynamics for the price
of router \( i \) would be 
\begin{equation}
\label{eq:sde1}
dS_{i}(t)=\alpha _{i}\big (\mu _{i}-S_{i}(t)\big )dt+\sigma _{i}dW_{i}(t)
\end{equation}
where \( W_{i}(t) \) is a Wiener process and the correlation between two processes
\( i \) and \( j \) is \( Corr[dW_{j}(t),dW_{j}(t)]=\rho _{i,j} \). Recall
that a Wiener process has independent normal distributed increments with mean
\( 0 \) and variance \( t-s \), i.e. \( W_{i}(t)-W(s_{i})\sim N[0,\sqrt{t-s}] \)
where \( t>s \). For improved readability, we omit the indices in \( S_{i}(t) \),
etc., when they are irrelevant for the understanding.

In (eq. \ref{eq:sde1}) the drift term (the \( dt \) term) detracts when \( S \)
is bigger than \( \mu  \), and adds when \( S \) is less than \( \mu  \).
The amount of the increase is determined by \( \alpha  \). The diffusion (the
\( dW \) term) is independent of \( S \).

When the simulation has run for sufficiently long time, the price \( S_{i}(t) \)
becomes independent of the starting state of the system, and reaches a stationary
distribution \( P(t) \) of \( S(t) \). Using the Fokker-Planck equation (see
Appendix A) gives 
\begin{equation}
\label{eq:distSde1}
P(s)=C_{0}e^{-\frac{1}{2}\big (\frac{s-\mu }{\sigma /\sqrt{2\alpha }}\big )^{2}}
\end{equation}
where \( C_{0}=(\pi \sigma ^{2}/\alpha )^{-\frac{1}{2}} \), which is the density
function for a normal distribution \( N[\mu ,\frac{\sigma }{\sqrt{2\alpha }}] \).
We note first that the normal distribution is non-zero in all of \( (-\infty ,\infty ) \),
meaning that \( S(t) \) can take on negative values, something that is not
possible in a market with Farmer's dynamics. Second, the normal distribution
has many good properties, such as that a weighted sum of normal distributed
variables is also normal distributed. If prices are far from zero, this dynamics
may therefore be a convenient approximation to the true distribution. Third,
the stationary distribution only depends on the ratio \( \frac{\sigma ^{2}}{\alpha } \)
and can therefore not distinguish between separate variations in \( \sigma ^{2} \)
and \( \alpha  \), which has to be done by other means.

The observations \( \hat{S}(i) \), \( i\in [1,...,L] \) of the process \( S \)
in its stationary state are regularly spaced with distance \( \Delta t \).
Since \( E[S]=\mu , \) we estimate \( \mu  \) with 
\[
\hat{\mu }=\hat{E}[S]=\frac{1}{L}\sum _{i=1}^{L}\hat{S}(i)\]
 We estimate \( \sigma  \) after noting that \( E[(dS)^{2}]=\sigma ^{2}dt+O(dt\, dW) \).
Therefore, for small \( \Delta t \), 
\begin{eqnarray*}
\hat{\sigma }^{2} & = & \frac{1}{\Delta t}\hat{E}[(dS)^{2}]\\
 & = & \frac{1}{\Delta t}\frac{1}{L-1}\sum ^{L-1}_{i=1}\Big (\hat{S}(i+1)-\hat{S}(i)\Big )^{2}
\end{eqnarray*}
Since \( Var[S]=\frac{\sigma ^{2}}{2\alpha } \), the unbiased estimate is 
\begin{eqnarray*}
\frac{\hat{\sigma }^{2}}{2\hat{\alpha }} & = & \frac{L}{L-1}\Big (\hat{E}[S^{2}]-\hat{E}[S]^{2}\Big )
\end{eqnarray*}
so the estimate of \( \alpha  \) is 
\[
\hat{\alpha }=\frac{1}{2\Delta t}\frac{\sum ^{L-1}_{i=1}\Big (\hat{S}(i+1)-\hat{S}(i)\Big )^{2}}{\Big (\sum ^{L}_{i=1}\hat{S}(i)^{2}-\big (\sum ^{L}_{i=1}\hat{S}(i)\big )^{2}\Big )}\]

\begin{figure*}[ht]
{\centering \begin{tabular}{cccc}
\begin{sideways}
\hspace{2cm} density
\end{sideways}&
\includegraphics{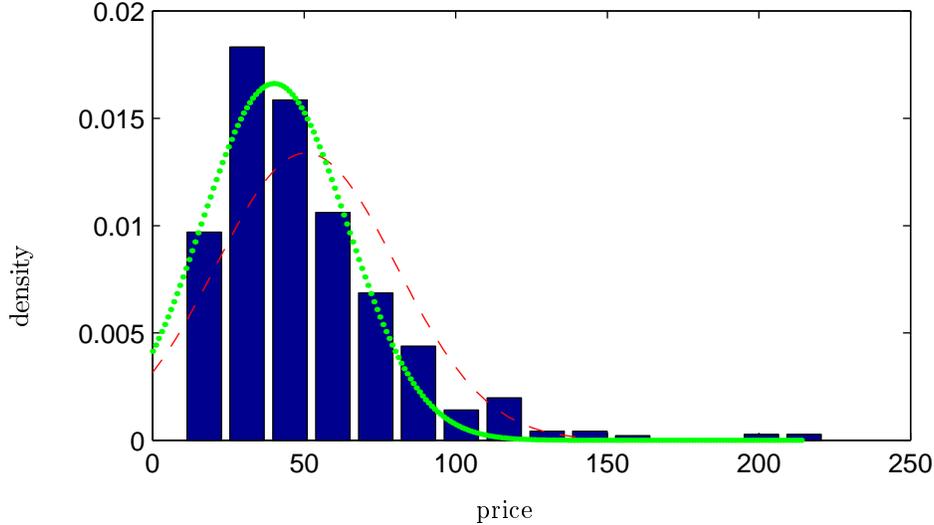} &
&
\\
&
price&
&
\\
\end{tabular}\par}

\caption{\label{fig:fit-sde1}Histogram over the observed prices, the estimated normal
distributed density function \protect\( N[\mu ,\sigma /\sqrt{2\alpha }]\protect \)
(the red dashed line), and a least square error fit to the 15 columns in the
histogram (the green dotted line).}
\end{figure*}
To verify that the observed estimate is indeed correct, we can plot the histogram
of the observed data together with the estimated density function \( P(s) \),
see (fig \ref{fig:fit-sde1}). This estimation coincides with the maximum likelihood
estimate of the variables. A least square fit to the values in the histogram
gives a better-looking curve. However, this fit only gives \( \mu  \) and the
ratio \( \frac{\sigma ^{2}}{\alpha } \), as noted above. 

Having estimates of \( \alpha  \), \( \mu  \) and \( \sigma  \) we are able
to estimate the correlation \( \rho _{i,j} \) between the random sources for
two price processes \( S_{i} \) and \( S_{j} \) by solving \( dW_{i} \) and
\( dW_{j} \) in (eq. \ref{eq:sde1}) and using the estimates
\[
d\hat{W}_{i}(k)=\frac{\hat{S}_{i}(k+1)-\hat{S}_{i}(k)-\hat{\alpha }_{i}\big (\hat{\mu }_{i}-\hat{S}_{i}(k)\big )\Delta t}{\hat{\sigma }_{i}}\]
 so that the estimate becomes 
\begin{eqnarray*}
\hat{\rho }_{i,j} & = & \frac{\hat{\textrm{C}}\textrm{ov}[dW_{i},dW_{j}]}{^{\hat{\textrm{V}}\textrm{ar}[dW_{i}]\hat{\textrm{V}}\textrm{ar}[dW_{j}]}}\\
 & = & \frac{\frac{1}{L-1}\sum ^{L-1}_{k=1}d\hat{W}_{i}(k)d\hat{W}_{j}(k)}{(\Delta t)^{2}}
\end{eqnarray*}
where we have used \( E[dW]=0 \) and \( Var[dW]=dt \).

\paragraph{Inhomogeneous drift, multiplicative noise, constant coefficients}

The dynamics in (sec. \ref{sec:sde1}) can generate negative prices, something
which is should not be possible in a well-functioning market. By asserting a
mean-reverting dynamics with multiplicative noise we get a process that is strictly
positive. Assume the dynamics for the price of router \( i \) to be 
\begin{equation}
\label{eq:sde2}
dS_{i}(t)=\alpha _{i}\big (\mu _{i}-S_{i}(t)\big )dt+\sigma _{i}S_{i}(t)dW_{i}(t)
\end{equation}
where \( W_{i}(t) \) is a Wiener process and the correlation between two processes
\( i \) and \( j \) is \( Corr[dW_{j}(t),dW_{j}(t)]=\rho _{i,j} \). As before,
we we omit the indices in \( S_{i}(t) \), etc., for readability.

The stationary distribution \( P(s) \) of \( S \) is 
\begin{equation}
\label{eq:distSde2}
P(s)=\frac{(\gamma \mu )^{\gamma }\mu }{\Gamma (\gamma )}e^{-\frac{\gamma \mu }{s}}\Big (\frac{1}{s}\Big )^{\gamma +2}
\end{equation}
where \( \gamma \equiv \frac{2\alpha }{\sigma ^{2}} \), and \( \Gamma (z) \)
is the gamma function. In (eq. \ref{eq:distSde2}) \( s \) takes only positive
values, which is consistent, since the dynamics in (eq. \ref{eq:sde2}) does
not move an \( s \) from positive to negative.

The first moment of this distribution, \( E[S] \) is \( \mu  \) (see Appendix
A). As for the Ornstein-Uhlenbeck process above, we estimate \( \mu  \) with
\[
\hat{\mu }=\hat{E}[S]=\frac{1}{L}\sum _{i=1}^{L}\hat{S}(i)\]
 We cannot easily estimate \( \sigma  \) from \( E[(dS)^{2}] \) since \( S \)
is in the drift term, but note that the process \( X(t)=\log (S(t)) \) in (eq.
\ref{eq:dX}) has additive noise. Therefore, we get for small \( \Delta t \),
\begin{eqnarray*}
\hat{\sigma }^{2} & = & \frac{1}{\Delta t}\hat{E}[(dX)^{2}]\\
 & = & \frac{1}{\Delta t}\frac{1}{L-1}\sum ^{L-1}_{i=1}\Big (\log \frac{\hat{S}(i+1)}{\hat{S}(i)}\Big )^{2}
\end{eqnarray*}
To estimate \( \alpha  \), we try two approaches. First we use the conditional
expectation \( E[S(t+\tau )|S(t)] \) for \( S \) in the stationary state of
the system. Taking the partial derivative w.r.t \( \tau  \) gives a first order
ODE with the solution 
\[
E[S(t+\tau )|S(t)]=e^{-\alpha \tau }(S(t)-\mu )+\mu \]
 showing that the expected value approaches \( \mu  \) exponentially with \( \tau  \)
and \( \alpha  \) determines the speed of the return. Rearrange to keep \( e^{-\alpha \tau } \)
(which is independent of \( S(t) \)) on one side, take the logarithm and take
the expected value of both sides. Let \( \tau =k\Delta t \). We now have an
estimate for \( \alpha  \), 
\[
exp(-\hat{\alpha }\, k\Delta t)=\frac{1}{L-k}\sum ^{L-k}_{i=1}\frac{\hat{S}(i+k)-\hat{\mu }}{\hat{S}(i)-\hat{\mu }}\]
\begin{figure*}[ht]
{\centering \begin{tabular}{cccc}
\begin{sideways}
\hspace{2cm}\( e^{-\alpha \, k\Delta t} \)
\end{sideways}&
\resizebox*{7cm}{!}{\includegraphics{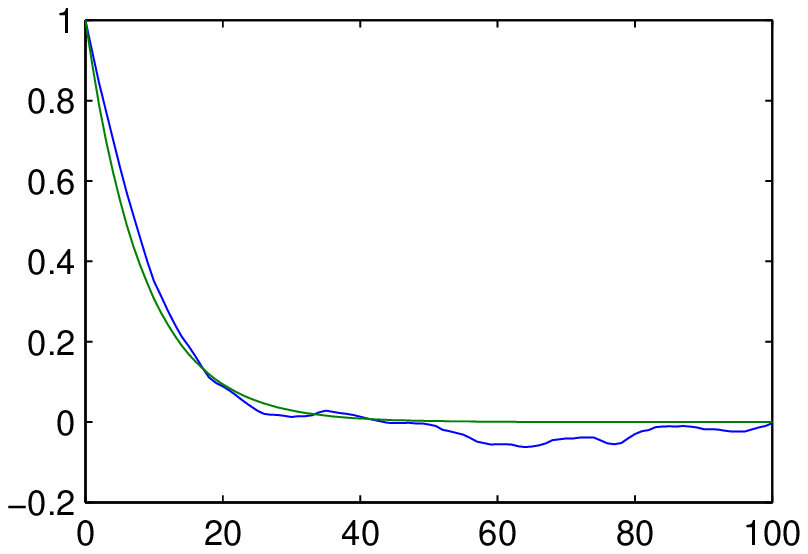}} &
\begin{sideways}
\hspace{1cm}\( \textrm{Cov}[S(t+k\Delta t),S(t)] \)
\end{sideways}&
\resizebox*{7cm}{!}{\includegraphics{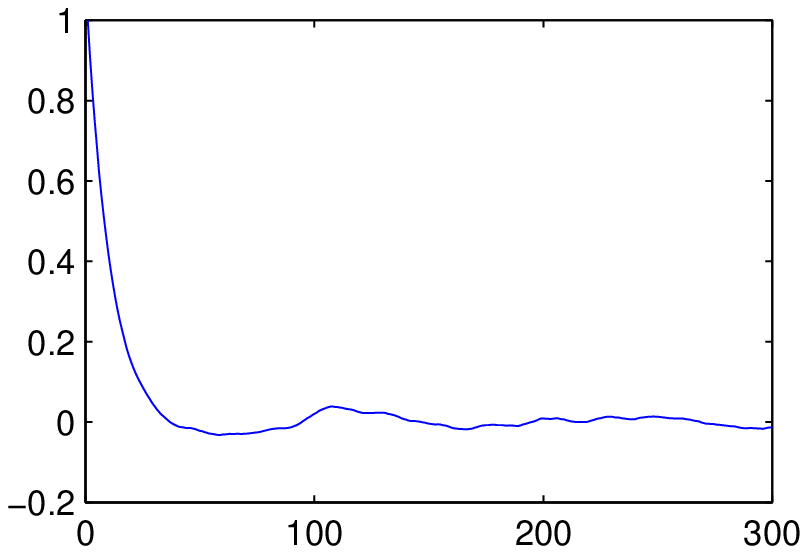}} \\

&
\( k=\tau /\Delta t \)&
&
\( k \)\\
\end{tabular}\par}

\caption{\label{fig:alphaestim}Left fig.: the estimate \protect\( \hat{y}=e^{-\alpha k\Delta t}\protect \)
derived from the conditional expectation of \protect\( S(t+\tau )\protect \)
for the first 20 time-steps in blue, and the plot of \protect\( y=e^{-\hat{\alpha }k\Delta t}\protect \)
in green. The estimate may be less than zero due to the noise from the simulation.
Right fig.: the covariance between \protect\( S(t+k\Delta t)\protect \) and
\protect\( S(t)\protect \). The auto-correlation deviates some from an exponential,
possibly because the price dynamics has higher order than assumed. It would
explain why the estimate in the left plot deviates from an exponential. Simulation
time, \protect\( 100\, 000\protect \) time steps, \protect\( \lambda _{i}=10,\protect \)
and \protect\( D=100\protect \).}
\end{figure*}
As we can see from the left hand side of the equation above, plotting the right
hand side as a function of \( k \) should result in a straight line if \( \alpha  \)
is constant. However, as we can see in (fig. \ref{fig:alphaestim}), the line
is straight up to some \( k \) that depends on the simulation parameters and
error. The estimation is very sensitive to errors in \( \hat{\mu } \), especially
in the denominator if \( \hat{S} \) is near \( \hat{\mu } \), which results
in a flattened out jagged line, since \( e^{-\hat{\alpha }\, k\Delta t} \)
is less than the simulation error for larger \( k \). It eventually flattens
out which means that \( \alpha  \) appears to decrease for larger \( \tau  \). 

\begin{figure*}[ht]
{\centering \begin{tabular}{cc}
\begin{sideways}
\hspace{2cm}density
\end{sideways}&
\includegraphics{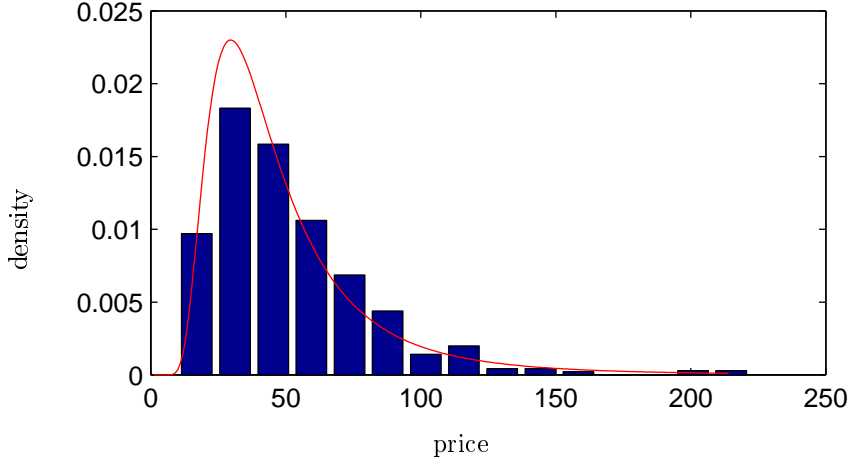} \\
&
price\\
\end{tabular}\par}

\caption{\label{fig:fit-sde2}Histogram over the observed prices and the estimated gamma
distribution. \protect\( \mu \protect \) and \protect\( \sigma \protect \)
are estimated using the moment method, and after that, \protect\( \alpha \protect \)
is estimated with a least square error fit. }
\end{figure*}
Another way to estimate \( \alpha  \) is to estimate \( \mu  \) and \( \sigma  \)
as above, and then fit the observed distribution to the distribution (eq. \ref{eq:distSde2})
using the least square method. See (fig. \ref{fig:fit-sde2}) for a plot of
the model fit.

Having estimates of \( \alpha  \), \( \mu  \) and \( \sigma  \) we are able
to estimate the correlation \( \rho _{i,j} \) between the random sources for
two price processes in the same way as above, by solving \( dW_{i} \) and \( dW_{j} \)
in (eq. \ref{eq:sde2}) and using the estimates
\[
\Delta \hat{W}_{i}(k)=\frac{\hat{S}_{i}(k+1)-\hat{S}_{i}(k)-\hat{\alpha }_{i}\big (\hat{\mu }_{i}-\hat{S}_{i}(k)\big )\Delta t}{\hat{\sigma }_{i}\hat{S}_{i}(k)}\]
 so that the estimate becomes 
\[
\hat{\rho }_{i,j}=\frac{\frac{1}{L-1}\sum ^{L-1}_{k=1}d\hat{W}_{i}(k)d\hat{W}_{j}(k)}{(\Delta t)^{2}}\]
where again we have used \( E[dW]=0 \) and \( Var[dW]=dt \). 

\begin{figure*}[ht]
{\centering \begin{tabular}{cclc}
\begin{sideways}
\hspace{2cm} density
\end{sideways}&
\includegraphics{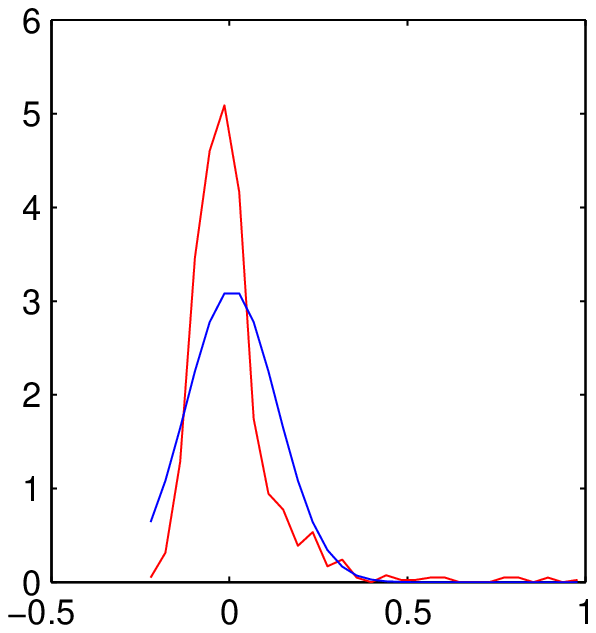} &
\begin{sideways}
\hspace{2cm} density
\end{sideways}&
\includegraphics{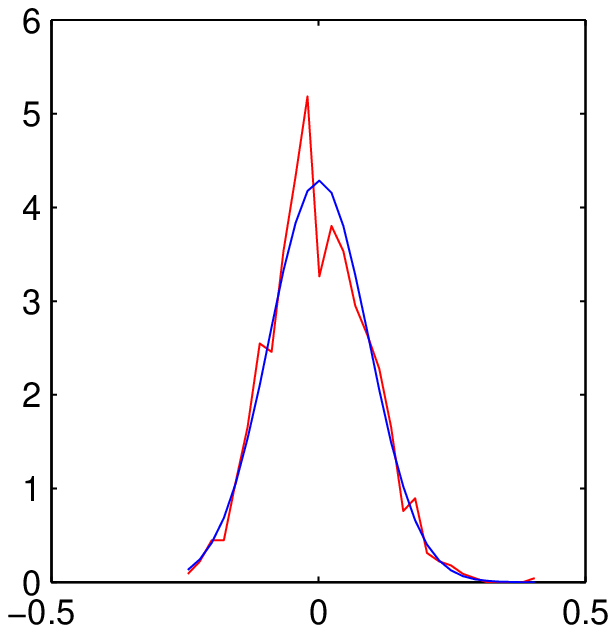} \\
&
\( \Delta \hat{W} \)&
&
\( \Delta \hat{W} \)\\
\end{tabular}\par}

\caption{\label{fig:fitdw}The left figure shows the histogram (in red) over \protect\( \Delta \hat{W}\protect \)
in the simulation described in (sec.\ref{sec:parameters}) with \protect\( \lambda _{i}=10\protect \),
and a fitted density function for a normal distribution (in blue). The right
figure shows the histogram of another simulation with the same parameters except
for the market liquidity which is \protect\( \lambda _{i}=100\protect \).}
\end{figure*}
Plotting the histograms of \( \Delta \hat{W} \) shows us if the model is good.
If that is the case, the \( \Delta \hat{W} \) should behave like samples of
a Wiener process, i.e. be normal distributed. As can be seen in (fig. \ref{fig:fitdw}),
the model (eq. \ref{eq:sde2}) fits well if the market liquidity is high. However,
if market liquidity is low then the assumed price model does not provide a good
fit.

\paragraph{Price Correlations}

The resource prices in a network of resources depend on the prices of other
resources, since they are traded in groups. Using the parameter estimates of
the price dynamics above, we get the correlation matrix in (tab. \ref{tab:corrlambda100})
for a simulation with high liquidity markets (\( \lambda _{i}=100) \).

\begin{table*}[ht]
{\centering \begin{tabular}{|c|c|c|c|c|c|c|c|c|c|c|}
\hline 
Router &
\textbf{1}&
\textbf{2}&
\textbf{3}&
\textbf{4}&
\textbf{5}&
\textbf{6}&
\textbf{7}&
\textbf{8}&
\textbf{9}&
\textbf{10}\\
\hline 
\hline 
\textbf{1}&
~~1~~&
\texttt{\textbf{\textit{0.33}}}&
-0.03&
0.06&
\textbf{\emph{0.48}}\emph{}&
\textbf{\emph{0.21}}&
-0.05&
0.08&
-0.04&
0.05\\
\hline 
\textbf{2}&
&
1&
0.02&
-0.12&
0.06&
-0.01&
0.04&
\textbf{\emph{0.60}}&
0.12&
\emph{0}\textbf{\emph{.36}}\\
\hline 
\textbf{3}&
&
&
1&
0.18&
0.02&
0.03&
\textbf{\emph{0.02}}&
0.14&
\textbf{\emph{0.43}}&
-0.03\\
\hline 
\textbf{4}&
&
&
&
1&
\textbf{\emph{0.38}}&
\textbf{\emph{0.40}}&
0.20&
-0.12&
\textbf{\emph{0.50}}&
0.09\\
\hline 
\textbf{5}&
&
&
&
&
1&
0.03&
0.05&
-0.09&
0.11&
-0.01\\
\hline 
\textbf{6}&
&
&
&
&
&
1&
0.02&
-0.11&
0.09&
\textbf{\emph{0.46}}\\
\hline 
\textbf{7}&
&
&
&
&
&
&
1&
0.15&
\textbf{\emph{0.45}}&
-0.02\\
\hline 
\textbf{8}&
&
&
&
&
&
&
&
1&
\textbf{\emph{0.41}}&
0.12\\
\hline 
\textbf{9}&
&
&
&
&
&
&
&
&
1&
0.01\\
\hline 
\textbf{10}&
&
&
&
&
&
&
&
&
&
1\\
\hline 
\end{tabular}\par}

\caption{\label{tab:corrlambda100}Correlation coefficients for \protect\( \Delta \hat{W}\protect \)
for a simulation with \protect\( \lambda _{i}=100\protect \), after 5000 time
steps. Connected routers are in bold face.}
\end{table*}
We find that prices generally are positively correlated. Comparing with the
network graph(fig. \ref{fig:network}), one can see that prices of neighboring
nodes often are strongly positively correlated with an average correlation coefficient
of around 0.4, compared to 0.03 for nodes that are not connected. The most significant
exception is nodes 3 and 7. Looking at the network graph it is clear that no
\emph{least cost path} can contain both 3 and 7, except for the path from 3
to 7.

\subsection{Efficiency of the market based resource allocation }

The efficiency of a market based resource allocation scheme depends on how well
prices reflect the available information about resource demands. Two sources
for bad performance is that price quotes are outdated, or that they do not reflect
knowledge about the price behavior, e.g. periodic price fluctuations, or future
prices. To measure efficiency, we can measure the resource utilization to compare
which markets are able to capture most information.

\paragraph{Communication costs}

The number of messages sent in the negotiation phase of this kind of system
is negligible, since all users operate on old price quotes and place bids at
market, i.e. accept the price whatever it may be. They do not update their price
quotes for every bid. Therefore, less than \( M \) messages per trade come
in to a user (the potential quote update from each market). One message per
trade (the bid) go out from each user to the markets that contain resources
that the user have chosen. No messages need to be communicated between the end-users.

\paragraph{Successful connection ratio}

\begin{figure*}[ht]
{\centering \begin{tabular}{cccc}
\begin{sideways}
\hspace{1cm} Successful connections 
\end{sideways}&
\resizebox*{7cm}{!}{\includegraphics{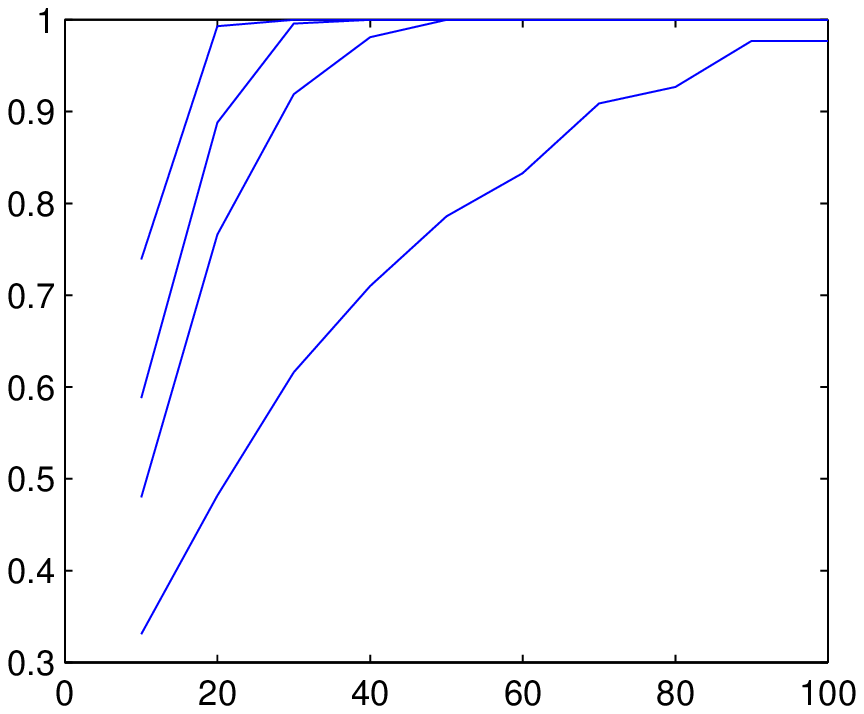}}  &
\begin{sideways}
\hspace{2cm} Avg. profit
\end{sideways}&
\resizebox*{7cm}{!}{\includegraphics{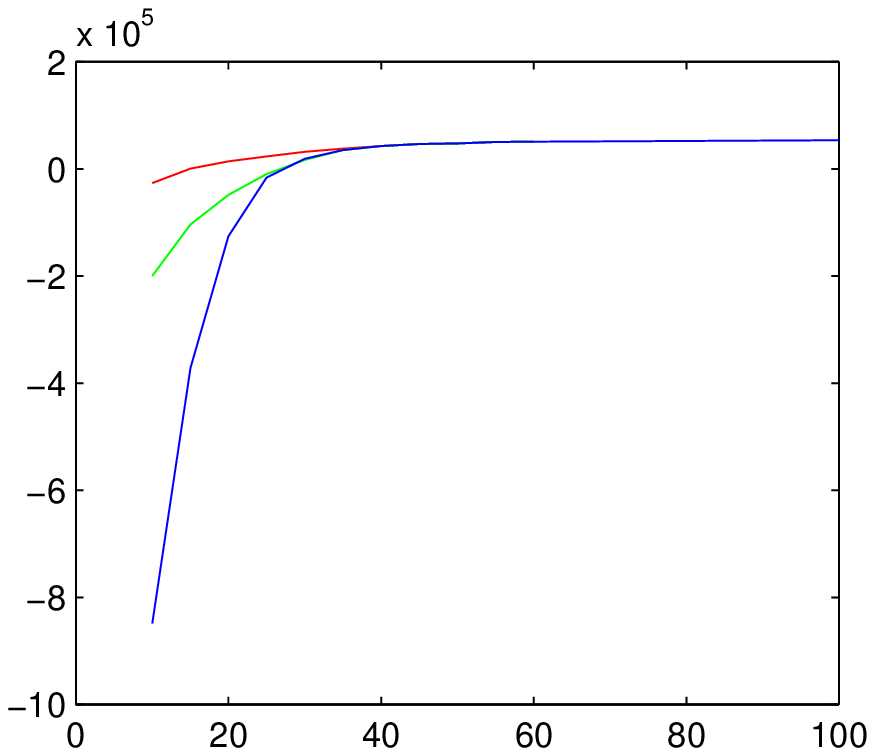}} \\
&
\( \lambda _{i} \)&
&
\( \lambda _{i} \)\\
\end{tabular}\par}

\caption{\label{fig:successfrac}To the left, average ratio of successful connections
as a function of the market liquidity. The graphs from bottom to top have \protect\( C_{max}\protect \)
equal to 1, 4, 16 and 64. To the right, average net profit as a function of
market liquidity. The graphs from bottom to top have \protect\( C_{max}\protect \)
equal to 64, 16, and 4, showing that a high \protect\( C_{max}\protect \) causes
losses on average in a low liquid market.}
\end{figure*}
In the Farmer market model, the less liquid the markets are, the less valid
are the price quotes. To the left in (fig. \ref{fig:successfrac}) the ratio
of successful connections is plotted as a function of the market liquidity \( \lambda _{i} \)
for simulations with the parameters described in (sec. \ref{sec:parameters}).
The graphs corresponds to different values of the decision parameter \( C_{max} \).
Low liquidity causes large price fluctuations, making the prices higher than
the limit \( C_{max}max \) (see sec. \ref{sec:setup}) which inhibits many
connections. Note that a connection is considered successful even if the net
cost (after releasing the resources) is higher than \( max \).

\paragraph{Net profit}

To the right in (fig. \ref{fig:successfrac}) we plot the average profit as
a function of the liquidity for a number of values of \( C_{max} \). Large
values of \( C_{max} \) causes the users to buy resources when they are expensive.
If the liquidity is high, the users sell resources at approximately the same
price, but with low liquidity, prices will move significantly downwards when
the resources are sold, causing a net loss to the user. 

The particular way the \( max \) cost for a connection is determined of course
very much determines which of the different kinds of traffic that is promoted
in the network. Different schemes could for instance promote short or long paths,
high or low capacity connections, etc.

\paragraph{Average load}

The average load, or reserved capacity in a router is with the Farmer dynamics
a direct function of the market price, \( \omega (t)=\lambda \log \frac{S(t)}{S(0)} \).
If two simulations that allocate the same connections differ in load, the higher
load depends on inefficient routing that does not choose efficient path. If
the simulations differ in allocated connections, the comparison is less straight
forward, and depends on the choice of metric above. We intend to return to a
longer discussion on efficiency, in particular for load balancing, in a future
contribution.

\section{Discussion\label{sec:Discussion}}

\label{sec:discussionFarmer}In our simulation, Farmer's market dynamics can
be interpreted either as prescriptions of how a rational market maker should
modify the market prices, or as a model of the aggregate behavior of market
price changes during some period of time. 

With the former interpretation, the Farmer dynamics can be used to implement
a market maker program that brokers trading between end-users. The dynamics
is derived from the assumption that it is impossible to change prices by 'trade
in circles'. This assumption is a necessary condition for any market maker strategy,
since otherwise anyone can exploit the market and gain an unlimited amount of
money from the market maker. 

The latter interpretation can be used when we want to simulate a part of a market
with many concurrent trades. Since trading causes prices to change means that
it is impossible to have updated price information at the time they place their
bid. It is only possible to have completely updated price info if the trading
is synchronous, something that severely reduces the number of bids a market
can handle per time unit if communication delays are taken into account.

The market liquidity (or market depth) parameter \( \lambda _{i} \) determines
the speed at which prices change. When the dynamics interpreted as prescriptions
for a market maker, it is up to the market maker to adjust \( \lambda _{i} \)
in order to reduce the risk of running out of resources. If we model an existing
bandwidth market, \( \lambda _{i} \) is an observable parameter which must
be determined from the distribution of price jumps in relation to traded volume.
The more resources in relation to the order size, the larger \( \lambda _{i} \).
A bandwidth market that is trading the capacity of a high capacity router can,
everything else being equal, be expected to have a higher \( \lambda _{i} \)
than a similar market for a congested or inefficient router.

In the mean reverting processes investigated here, \( \alpha  \) determines
the speed with which the process returns towards its statistical average \( \mu  \).
The speed of return depends on the characteristic time length of the system,
determined by the simulation parameter \( D \), the call length. With a large
\( D \), the prices will be correlated with previous prices over a longer time
period, which results in a small \( \alpha  \). With a small \( D \), the
price effect of previous calls will soon be forgotten, resulting in a high \( \alpha  \). 

The traditional statistical models used for telephony have been found to inadequately
model data network traffic. Data communication has been found to show a very
bursty or fractal behavior as one communication event often generates a burst
of more communication to other parts of the network. Data communication is generally
short lived and often with strong latency bounds, due to the increasing use
of computer networks for interactive communication. For short lived communication
with low transfered amount, traditional switched best effort networks will probably
continue to provide a very efficient solution for a long time. However, with
increasing demand of streaming real-time data such as high quality video-telephony,
video-on-demand, etc., it is necessary to be able to reserve capacity. The alternatives
are large buffers, which has bad latency performance, or migrating data (intelligent
replication), which can reduce load for one-to-many communication. Neither is
suited for point-to-point communication service guarantees.

In the 'bandwidth market' presented in this paper, end users buy the required
resources themselves. The demand for updated price quotes results generates
additional network traffic. In an extended model, we could allow risk neutral
middle men to sell options on the resources. Since updated quotes will then
only be required by the (few) middle men doing the actual trading, the additional
traffic would be greatly reduced. We intend to return to these topics in future
work.

\section*{Acknowledgement}

This work was supported by the Swedish National Board for Industrial and Technical
Development (NUTEK), project P11253, COORD, in the Promodis program, and by
the Swedish Research Institute for Information Technology (SITI), project eMarkets,
in the Internet3 program. We want to thank Sverker Janson and Bengt Ahlgren
for helpful discussions and comments.

\section*{Appendix A}

\subsection*{Stationary Distribution}

\( S(t) \) is an It\^o process with the dynamics \( dS(t)=a(t,S(t))dt+b(t,S(t))dW(t) \),
and \( P(s,T;s_{0},t) \) is the contingent probability distribution of \( S \)
at time \( T \) given \( S(t)=s_{0} \). \( P \) obeys the Fokker-Planck equation
a.k.a the forward Kolmogorov equation,
\begin{equation}
\label{eq:fokkerplanck}
-\partial _{T}P-\partial _{s}\Big (a(T,s)\, P\Big )+\partial _{ss}\big (\frac{b(T,s)^{2}}{2}P\big )=0
\end{equation}
 An important class of stochastic processes are those that are stationary. For
those, if \( T-t \) is sufficiently large, the contingent distribution \( P \)
no longer depends on \( s_{0} \), and can be substituted with the stationary
distribution \( P(s) \), which does not depend on \( s_{0} \), \( t \) or
\( T \). Two ways in which a stochastic process can fail to be stationary is
if the coefficients, e.g. \( a \) and \( b \), are explicitly time-dependent,
or if there is no distribution \( P(s) \). The latter happens for instance
in ordinary random diffusion, in which the probability gradually spreads out
without reaching a limit. 

Assuming that a stationary distribution has been reached, the time derivative
drops out of (eq. \ref{eq:fokkerplanck}), and the equation can be integrated
once to be
\begin{equation}
\label{eq:ODE}
\partial _{s}P(s)-\frac{2a(s)-2b(s)\partial _{s}b(s)}{b(s)^{2}}P(s)=0
\end{equation}
 which is an ODE. One further integration of (eq. \ref{eq:ODE}) gives
\begin{equation}
\label{eq:statDens}
P(s)=C\, e^{\int _{0}^{s}\frac{2a(u)}{b(u)^{2}}du}\frac{1}{b(s)^{2}}
\end{equation}
 where \( C \) is a normalization constant determined by \( \int P(s)ds=1 \).

\subsection*{The Additive Noise Process}

\paragraph{Stationary density function, mean and variance}

Assume the price dynamics 
\[
dS(t)=\alpha (\mu -S(t))dt+\sigma dW(t)\]
 Then the conditional stationary probability distribution of \( S \) is, using
(eq. \ref{eq:statDens}), 
\begin{eqnarray*}
P(s) & = & C_{0}e^{-\frac{2\alpha }{\sigma ^{2}}\int _{0}^{s}(\mu -u)du}\\
 & = & C_{0}e^{-\frac{1}{2}\big (\frac{s-\mu }{\sigma /\sqrt{2\alpha }}\big )^{2}}
\end{eqnarray*}
where \( C_{0}=(\pi \sigma ^{2}/\alpha )^{-\frac{1}{2}} \). \( P(s) \) can
be identified as the density function for a normal distribution with mean \( \mu  \)
and variance \( \sigma ^{2}/2\alpha  \).

\subsection*{The Multiplicative Noise Process}

\paragraph{Stationary density function }

Assume the price dynamics 
\[
dS(t)=\alpha (\mu -S(t))dt+\sigma SdW(t)\]
 Using (eq. \ref{eq:statDens})
\begin{eqnarray*}
P(s) & = & C_{1}e^{\frac{2\alpha }{\sigma ^{2}}\big (\int _{s_{0}}^{s}\frac{\mu }{u^{2}}du-\overbrace{\int _{s_{0}}^{s}\frac{1}{u}du}^{=\log s-\log s_{0}}\big )}\Big (\frac{1}{\sigma \, s}\Big )^{2}\\
 & = & \frac{1}{C_{2}}e^{\frac{2\alpha }{\sigma ^{2}}\big (-\frac{\mu }{s}\big )}\Big (\frac{1}{s}\Big )^{\frac{2\alpha }{\sigma ^{2}}+2}
\end{eqnarray*}

An alternative derivation is to let \( X(t)=\log S(t) \). It\^o's lemma gives
that
\begin{equation}
\label{eq:dX}
dX=\Big (\alpha \mu e^{-X}-(\alpha +\frac{\sigma ^{2}}{2})\Big )dt+\sigma dW
\end{equation}
 Then the stationary probability distribution \( Q(x) \) of \( X \) is , using
(eq. \ref{eq:statDens}),
\begin{eqnarray*}
Q(x) & = & C_{0}e^{\int _{0}^{x}\big (-\frac{2\big (\alpha \mu e^{-X}-(\alpha +\frac{\sigma ^{2}}{2})\big )}{\sigma ^{2}}\big )du}\frac{1}{\sigma ^{2}}\\
 & = & C_{1}e^{-\gamma \mu e^{-x}+(\gamma +1)x}
\end{eqnarray*}
 where \( C_{1} \) is a normalization constant, and \( \gamma \equiv \frac{2\alpha }{\sigma ^{2}} \)
for readability. The stationary density function \( P(s) \) for \( S \) is
found by recalling \( X=\log (S) \) and 

\begin{eqnarray*}
P(s)ds & \equiv  & Q(\log (s))d(\log (s))\\
 & = & \frac{1}{C_{2}}e^{-\gamma \mu \frac{1}{s}}\Big (\frac{1}{s}\Big )^{\gamma +1}\frac{1}{s}ds
\end{eqnarray*}

The constant \( C_{2} \) is determined by normalization and identifying the
integral as a gamma function, \( \Gamma (z+1)=\int ^{\infty }_{0}t^{z}e^{-t}dt=z\Gamma (z) \).
After the substitution \( t=\gamma \mu \frac{1}{s} \) we get
\begin{eqnarray*}
C_{2} & = & \int ^{0}_{\infty }e^{-t}\Big (\frac{1}{\gamma \mu }t\Big )^{\gamma +2}(-\gamma \mu \frac{1}{t^{2}})dt\\
 & = & \Big (\frac{1}{\gamma \mu }\Big )^{\gamma +1}\int ^{\infty }_{0}e^{-t}t^{\gamma }dt\\
 & = & \Big (\frac{1}{\gamma \mu }\Big )^{\gamma }\frac{1}{\mu }\Gamma (\gamma )
\end{eqnarray*}

\paragraph{Stationary mean}

Use the same substitution, \( t=\gamma \mu \frac{1}{s} \) to obtain 

\begin{eqnarray*}
E[S] & = & \int ^{\infty }_{0}sP(s)ds\\
 & = & \int ^{0}_{\infty }(\frac{1}{\gamma \mu }t)\frac{1}{C_{2}}e^{-t}\Big (\frac{1}{\gamma \mu }t\Big )^{\gamma +2}(-\gamma \mu \frac{1}{t^{2}})dt\\
 & = & \frac{1}{C_{2}}\Big (\frac{1}{\gamma \mu }\Big )^{\gamma }\underbrace{\int ^{\infty }_{0}e^{-t}t^{\gamma -1}dt}_{=\Gamma (\gamma )}\\
 & = & \mu 
\end{eqnarray*}

\paragraph{Conditional expected value }

Let \( f(\tau |S(t)) \) be the conditional expectation of \( S(t+\tau ) \)
given \( S(t) \).

\begin{eqnarray*}
f(\tau |S(t)) & \equiv  & E[S(t+\tau )|S(t)]\\
 & = & S(t)+E[\int _{t}^{t+\tau }dS(t')|S(t)]\\
 & = & S(t)+\alpha \mu \tau -\alpha \int _{t}^{t+\tau }E[S(t')|S(t)]dt'
\end{eqnarray*}
 Take the partial derivative with respect to \( \tau  \) 
\begin{eqnarray*}
\partial _{\tau }f(\tau |S(t)) & = & \alpha \mu -\alpha E[S(\tau )|S(t)]\\
 & = & \alpha \mu -\alpha f(\tau |(S(t))
\end{eqnarray*}
 Multiplying with \( e^{\alpha \tau } \) and collecting the terms gives 
\[
\partial _{\tau }\Big (e^{\alpha \tau }f(\tau |(S(t))\Big )=e^{\alpha \tau }\alpha \mu \]
Integrate both sides over \( \tau  \) from \( 0 \) to \( \tau  \) and multiplication
with \( e^{-\alpha \tau } \) gives 
\[
f(\tau |S(t))=e^{-\alpha \tau }\big (S(t)-\mu \big )+\mu \]
 where we have used that \( f(0|S(t))=S(t) \). Therefore 
\[
e^{-\alpha \tau }=\frac{f(\tau |S(t))-\mu }{S(t)-\mu }\]
The left side does not depend on \( S(t) \). Taking the expected value of both
sides
\begin{eqnarray*}
e^{-\alpha \tau } & = & E[e^{-\alpha \tau }]\\
 & = & E\Big [\frac{E[S(t+\tau )|S(t)]-\mu }{S(t)-\mu }\Big ]\\
 & = & \int _{0}^{\infty }\frac{E[S(t+\tau )|S(t)=x]-\mu }{x-\mu }P(x)dx
\end{eqnarray*}
where \( P \) is the stationary distribution of \( S \), and we have substituted
a time average with an ensemble average, which is true under suitable assumptions.

\paragraph{Estimation of parameter \protect\( \alpha \protect \) }

We can estimate the integral from the observed data using the estimates 
\begin{eqnarray*}
\hat{P}(x) & \equiv  & \frac{1}{N}\sum _{t}\delta (x-\hat{S}(t))\\
\hat{E}[S(t+\tau )|S(t)=x] & \equiv  & \frac{\sum _{t|\hat{S}(t)=x}\hat{S}(t+\tau )}{|\{x|\hat{S}(t)=x\}|}
\end{eqnarray*}
 where \( \delta (x) \) is the Dirac delta function and \( |A| \) is the cardinality
of the set \( A \). The estimation is 
\begin{eqnarray*}
e^{-\hat{\alpha }\tau } & \equiv  & \int _{0}^{\infty }\hat{P}(x)\frac{\hat{E}[S(t+\tau )|S(t)=x]-\hat{\mu }}{x-\hat{\mu }}dx\\
 & = & \int _{0}^{\infty }\Big (\frac{1}{N}\sum _{t}\delta (x-\hat{S}(t))\Big )\hat{E}[\frac{S(t+\tau )-\hat{\mu }}{x-\hat{\mu }}|S(t)=x]dx\\
 & = & \sum _{x|\hat{S}(t)=x}\frac{1}{N}|\{x|\hat{S}(t)=x\}|\frac{\sum _{t|\hat{S}(t)=x}\Big (\frac{S(t+\tau )-\hat{\mu }}{x-\hat{\mu }}\Big )}{|\{x|\hat{S}(t)=x\}|}\nonumber \\
 & = & \frac{1}{N}\sum _{t}\frac{\hat{S}(t+\tau )-\hat{\mu }}{\hat{S}(t)-\hat{\mu }}
\end{eqnarray*}
 so we can estimate \( \alpha  \) with 
\[
\hat{\alpha }(\tau )\equiv -\frac{1}{\tau }\log \Big (\frac{1}{N}\sum _{t}\frac{\hat{S}(t+\tau )-\hat{\mu }}{\hat{S}(t)-\hat{\mu }}\Big )\]
which should be a constant function.

\subsection*{Farmer's market dynamics}

Let \( S \) be the current price, \( \omega  \) the net demand and \( \tilde{S} \)
the price at which the demand can be met. We seek a functional relationship
of the type \( \tilde{S}\equiv \tilde{S}(S,\omega ) \), where \( \tilde{S} \)
depends continuously on \( S \) and \( \omega  \). Assume the price to be
positive and bounded, \( \tilde{S} \) to be an increasing function of \( \omega  \),
and that prices are only changed through trading, \( \tilde{S}(S,0)=S \). Assume
furthermore that one cannot make money by trading in circles 
\begin{equation}
\label{farmercirkel}
\tilde{S}(\tilde{S}(\tilde{S}(S,\omega _{1}),\omega _{2}),-(\omega _{1}+\omega _{2}))=S
\end{equation}
 and the relative price change is independent of the absolute price,
\begin{equation}
\label{farmer:relprice}
\frac{\tilde{S}(S,\omega )}{S}=\phi (\omega )
\end{equation}
 Because of (\ref{farmercirkel}) we see that \( \tilde{S}(\tilde{S}(S,\omega ),-\omega )=S \),
so the inverse is \( \tilde{S}^{-1}(S,\omega )=\tilde{S}(S,-\omega ) \), where
\( \tilde{S}^{-1} \) is the inverse function of \( \tilde{S} \). Applying
\( \tilde{S}^{-1}(\cdot ,-(\omega _{1}+\omega _{2})) \) on (\ref{farmercirkel})
to see that
\begin{equation}
\label{farmer4.3}
\tilde{S}(\tilde{S}(S,\omega _{1}),\omega _{2})=\tilde{S}(S,\omega _{1}+\omega _{2})
\end{equation}
 (\ref{farmer:relprice}) together with (\ref{farmer4.3}) implies that \( \phi (\omega _{1})\phi (\omega _{2})=\phi (\omega _{1}+\omega _{2}) \)
which implies \( \phi (x)=e^{x/\lambda } \) for some constant \( \lambda  \).
Therefore
\begin{equation}
\label{farmer price}
\tilde{S}(S,\omega )=S\, e^{\omega /\lambda }
\end{equation}


\begin{thebibliography}{Waldspurger92}
\bibitem[Avellaneda99]{Avellaneda99}Marco Avellaneda and Peter Laurence, \emph{Quantitative Modeling of Derivative
Securities: From Theory to Practice,} CRC Press, 1999.
\bibitem[Black73]{Black73}Fischer Black and Myron Scholes, \emph{The Pricing of Options and Corporate
Liabilities,} Journal of Political Economy, \textbf{(81:3)}, pp. 637-654, 1973.
\bibitem[Faratin00]{Faratin00}P. Faratin, N. R. Jennings, P. Buckle and C. Sierra. \emph{Automated Negotiation
for Provisioning Virtual Private Networks using FIPA-Compliant Agents}, Proc.
of the 5th Int. Conf. on Practical Application of Intelligent Agents and Multi-Agent
Systems (PAAM-2000), Manchester, UK. 2000.
\bibitem[Farmer00]{Farmer00}J. Doyne Farmer, \emph{Market force, ecology, and Evolution,} submitted to Journal
of Economic Behavior and Organization, Feb. 2000.
\bibitem[Ferguson88]{Ferguson88}Donald Ferguson, Yechiam Yemini, and Christos Nikolaou, \emph{Microeconomic
Algorithms for Load Balancing in Distributed Computer Systems}, Proc. of the
8th Int. Conf. on Distributed Computer System, 1988.
\bibitem[Kurose85]{Kurose85}James F. Kurose, Mischa Schwartz, and Yechiam Yemini, \emph{A Microeconomic
Approach to Optim}ization of Channel Access Policies in Multi-Access Networks,
Proc. of 5th Int. Conf. on Distributed Computer Systems, 1985.
\bibitem[Kurose89]{Kurose89}James F. Kurose, and Rahul Simha, \emph{A Microeconomic Approach to Optimal
Resource Allocation in Distributed Computer Systems}, IEEE Trans. on Computers,
\textbf{(38)} no. 5, May 1989.
\bibitem[Lazar98]{Lazar98}Aurel A. Lazar, and Nemo Semret, \emph{Spot and Derivative Markets for Admission
Control and Pricing in Connection-Oriented Networks,} CU/CTR/TR 501-98-35, Columbia
University, New York, \emph{1998.}
\bibitem[Rassenti82]{Rassenti82}S. J. Rassenti, V. L. Smith, and R. L. Buffin, \emph{A Combinatorial Auction
Mechanism for Airport Time Slot Allocation}, The Bell Journal of Economics,
\textbf{(13)} pp. 402-417,1982.
\bibitem[Rothkopf98]{Rothkopf98}Michael G, Rothkopf, Aleksandar Peke\v c, and Ronald M. Harstad, \emph{Computationally
Manageable Combinational Auctions,} Management Science, \textbf{(44)} no. 8,
Aug. 1998.
\bibitem[Sairamesh95]{Sairamesh95}Jakka Sairamesh, Donald F. Ferguson, and Yechiam Yemini, \emph{An Approach to
Pricing, Optimal Allocation and Quality of Service Provisioning In High-Speed
Packet Networks, IEEE INFOCOM, 1995.}
\bibitem[Sandholm99]{Sandholm99}Tuomas Sandholm, \emph{An Algorithm for Optimal Winner Determination in Combinatorial
Auctions,} Proc. of 16th Int. Joint Conf. on Artificial Intelligence (IJCAI'99),
July, 1999.
\bibitem[Waldspurger92]{Waldspurger92}Carl A. Waldspurger, Tad Hogg, Bernardo A. Huberman, Jeffrey O. Kephart, and
W. Scott Stornetta, \emph{Spawn: A Distributed Computational Economy}, IEEE
Trans. on Software Eng. \textbf{(18)} no. 2, Feb. 1992.
\end{thebibliography}
\end{document}